\documentclass[fleqn,usenatbib]{mnras}

\usepackage{newtxtext, newtxmath, nicefrac}
\usepackage[T1]{fontenc}

\DeclareRobustCommand{\VAN}[3]{#2}
\let\VANthebibliography\thebibliography
\def\thebibliography{\DeclareRobustCommand{\VAN}[3]{##3}\VANthebibliography}

\usepackage{graphicx}
\usepackage{amsmath}

\newcommand{\txa}{{\text{a}}}

\newcommand{\txB}{{\text{B}}}

\newcommand{\txd}{{\text{d}}}
\newcommand{\txe}{{\text{e}}}

\newcommand{\txt}{{\text{t}}}
\newcommand{\calE}{{\cal{E}}}

\newcommand{\rT}{r_{\scriptscriptstyle{\text{T}}}}

\newcommand{\ra}{r_\txa}
\newcommand{\bfx}{\boldsymbol{x}}
\newcommand{\bfv}{\boldsymbol{v}}

\DeclareMathOperator{\arctanh}{arctanh}
\DeclareMathOperator{\erf}{erf}
\DeclareMathOperator{\floor}{floor}

\defcitealias{2022MNRAS.512.2266B}{Paper~I}
\defcitealias{2023MNRAS.519.6065B}{Paper~II}

\title[Dynamical models with a finite extent~III]{Self-consistent dynamical models with a finite extent -- III. Truncated power-law spheres}

\author[M. Baes \& B. Vander Meulen]{Maarten Baes and Bert Vander Meulen
\\%
Sterrenkundig Observatorium, Universiteit Gent, Krijgslaan 281 S9, 9000 Gent, Belgium
}

\date{Accepted 2023 July 20. Received 2023 July 6; in original form 2023 May 24}

\pubyear{2023}

\begin{document}
\label{firstpage}
\pagerange{\pageref{firstpage}--\pageref{lastpage}}
\maketitle

\begin{abstract}
Fully analytical dynamical models usually have an infinite extent, while real star clusters, galaxies, and dark matter haloes have a finite extent. The standard method for generating dynamical models with a finite extent consists of taking a model with an infinite extent and applying a truncation in binding energy. This method, however, cannot be used to generate models with a pre-set analytical mass density profile. We investigate the self-consistency and dynamical properties of a family of power-law spheres with a general tangential Cuddeford (TC) orbital structure. By varying the density power-law slope $\gamma$ and the central anisotropy $\beta_0$, these models cover a wide parameter space in density and anisotropy profiles. We explicitly calculate the phase-space distribution function for various parameter combinations, and interpret our results in terms of the energy distribution of bound orbits. We find that truncated power-law spheres can be supported by a TC orbital structure if and only if $\gamma \geqslant 2\beta_0$, which means that the central density slope--anisotropy inequality is both a sufficient and a necessary condition for this family. We provide closed expressions for structural and dynamical properties such as the radial and tangential velocity dispersion profiles, which can be compared against more complex numerical modelling results. This work significantly adds to the available suite of self-consistent dynamical models with a finite extent and an analytical description.
\end{abstract}



\section{Introduction}

In spite of the ever increasing power of computers and advances in modelling techniques, fully analytical dynamical models remain an important tool for the study of dynamical structures such as galaxies or dark matter haloes. They are often the starting point for more complex numerical modelling and can serve as a representative environment in which new modelling or analysis techniques can be tested, or in which the effects of additional physical processes can be investigated. Some of the most popular textbook models include the Plummer sphere \citep{1911MNRAS..71..460P}, the Hernquist model \citep{1990ApJ...356..359H}, and the isochrone sphere \citep{1959AnAp...22..126H}, and more generally, the families of $\gamma$--models \citep{1993MNRAS.265..250D, 1994AJ....107..634T}, Einasto models \citep{1965TrAlm...5...87E}, or generalised NFW models  \citep{2000ApJ...529L..69J, 2006AJ....132.2685M}. Many photometric, dynamical, and lensing properties of these models can be calculated analytically, and they have been explored in depth over the past decades. For example, for the Plummer model alone, the list of studies exploring the dynamical characteristics using analytical techniques is impressive \citep[e.g.,][]{1979PAZh....5...77O, 1985AJ.....90.1027M, 1986PhR...133..217D, 1987MNRAS.224...13D, 1991MNRAS.253..414C, 1996A&A...312..649W, 2007A&A...471..419B, 2013PhRvD..88f4020N, 2014MNRAS.440.2636L, 2019MNRAS.487..711R}.

One common aspect of nearly all of these popular models is that they have an infinite extent, meaning that their density is positive and non-zero over the entire space. Simple models with an analytical density profile with a finite extent would be very useful, however, as real dynamical systems such as star clusters or galaxies have a finite extent. Moreover, when setting up numerical experiments to test a new modelling technique (e.g., a new orbit integrator), analytical models with a finite extent form a convenient starting point.

The standard approach for generating dynamical models with a finite extent consists of taking a model with an infinite extent as a starting point and applying a truncation in binding energy. This approach is supported by various studies arguing that tidal stripping may be best described as a truncation process in binding energy space \citep{2009MNRAS.400.1247C, 2017MNRAS.468.2345D, 2020MNRAS.494..378D, 2021arXiv211101148A, 2021MNRAS.508.5196S, 2023MNRAS.521.4432S}. The standard example of models generated in this way is the family of King models \citep{1963MNRAS.125..127M, 1966AJ.....71...64K}, which are constructed by applying a binding energy truncation to the isothermal sphere. Several other models with an energy truncation have been proposed \citep[e.g.,][]{1954MNRAS.114..191W, 1970AJ.....75..674P, 1975AJ.....80..175W, 2014JSMTE..04..006G, 2017MNRAS.468.2345D}. A very versatile family of models belonging to this class is the family of lowered isothermal of {\sc{limepy}} models \citep{2015MNRAS.454..576G}, which have been shown to provide good fits to both simulated and observed globular clusters \citep{2016MNRAS.462..696Z, 2023MNRAS.519..445C}. The main disadvantage of all of these energy-truncated models is that, while the distribution function can be expressed analytically in terms of the integrals of motion, even the most simple radial profiles such as the density profile and gravitational potential cannot be expressed exactly. In addition, many of these models exclude orbits in an allowed part of phase space to achieve their radial truncation \citep{1988ApJ...325..566K}.

We have started an investigation into the possibility to construct simple dynamical models for spherical systems with a preset density profile with a finite extent. The overall aims are to investigate whether it is possible to build self-consistent dynamical models corresponding to such density profiles and if so, to find out which orbital structure would support them. In \citet{2022MNRAS.512.2266B}, hereafter \citetalias{2022MNRAS.512.2266B}, we started this endeavour by focusing on the uniform density sphere, the simplest model with a radially truncated density profile. We demonstrated that the uniform density sphere cannot be supported by an ergodic, constant anisotropy, or radial Osipkov--Merritt orbital structure, but we constructed a family of fully analytical dynamical models that could support this model. In \citet{2023MNRAS.519.6065B}, hereafter \citetalias{2023MNRAS.519.6065B}, we investigated in a systematic way which orbital structures could support radially truncated models, and formulated a consistency hypothesis. 

In this paper, the third one in this series, we present a detailed analysis of the dynamical properties of an interesting family of finite models: the truncated power-law spheres. Spherical models in which the density is a pure power law over the entire radial range, $\rho(r) \propto r^{-\gamma}$ with $\gamma\geqslant0$, have the advantage that many structural and dynamical properties can be calculated analytically \citep{1994MNRAS.267..333E, 2008gady.book.....B, 2021isd..book.....C}. The special case $\gamma=2$, known as the singular isothermal sphere (SIS), is probably the most famous member of the class of power-law spheres and is widely used in stellar dynamics and lensing studies \citep[e.g.,][]{1991ApJ...373..354K, 1993ApJ...419...12K, 1994ApJ...436...56K, 1991MNRAS.253...99F, 1991MNRAS.250..812G, 1993MNRAS.265..213G, 2009MNRAS.393..491C, 2014MNRAS.443..328L}. On the other hand, power-law spheres without radial truncation have the disadvantage that they always have an infinite total mass. Indeed, the total mass profile diverges in the centre if $\gamma\geqslant 3$ and at large radii if $ \gamma\leqslant 3$. 

If we truncate the density of the family of power-law spheres at a given radius, the truncation radius, we obtain a family of models with a finite extent, and with a finite total mass if $\gamma<3$. There are two important reasons why this family of truncated power-law spheres deserve to be the topic of an in-depth investigation. Firstly, this family covers a very wide range of density profiles, ranging from a point mass for $\gamma\to3$ to the uniform density sphere at $\gamma=0$, and even to models with a central density hole for $\gamma<0$. Secondly, the density profile of truncated power-law spheres is simple enough that most of the interesting dynamical characteristics can be evaluated completely analytically, for different choices of the orbital structure. 

As discussed in \citetalias{2023MNRAS.519.6065B}, truncated models with a density discontinuity, such as the truncated power-law spheres, can never be supported by an ergodic orbital structure. A promising alternative for truncated density models is the tangential Osipkov--Merritt (TOM) orbital structure \citep{1985AJ.....90.1027M}. The TOM orbital structure is characterised by an anisotropy profile
\begin{equation}
\beta(r) = - \frac{r^2}{\ra^2-r^2},
\end{equation}
with $\ra$ is the so-called anisotropy radius. In \citetalias{2023MNRAS.519.6065B} we formulated a consistency hypothesis, which states that for any non-truncated model that can be supported by an ergodic orbital structure, the corresponding truncated density profile with truncation radius $\rT$ can be supported by the TOM orbital structure with $\ra=\rT$. Based on this consistency hypothesis, one would expect that the truncated power-law spheres with $0\leqslant\gamma<3$ can be supported by the TOM orbital structure. Indeed, a truncated power-law sphere can be considered as a truncated version of a $\gamma$--model with the same value of $\gamma$ and with a scaling radius much larger than the truncation radius. Since all $\gamma$--models have positive ergodic distribution functions \citep{1993MNRAS.265..250D, 1994AJ....107..634T}, we expect the truncated power-law spheres with a TOM orbital structure to be consistent.

In this paper, we go one step further, however, and consider truncated power-law spheres with a tangential Cuddeford (TC) orbital structure. The TC orbital structure (\citealt{1991MNRAS.253..414C}; \citetalias{2022MNRAS.512.2266B}) is a generalisation of the TOM orbital structure, and is generally characterised by an anisotropy profile
\begin{equation}
\beta(r) = \beta_0 - (1-\beta_0)\left(\frac{r^2}{\ra^2-r^2}\right),
\label{TC-beta}
\end{equation}
with $\beta_0<1$, the central anisotropy, an additional free parameter. We consider truncated power-law spheres with a TC orbital structure with arbitrary values for $\beta_0$ and with $\ra=\rT$ to assure a completely tangential orbital structure at the truncation radius, as required for finite dynamical models (see \citetalias{2022MNRAS.512.2266B}). The detailed study of the resulting two-parameter family of dynamical models, with parameters $\gamma$ and $\beta_0$, is the topic of this third paper in this series. We note that, contrary to \citetalias{2023MNRAS.519.6065B} where we considered spherical models with an arbitrary density profile, we now focus on a specific family of spherical models with truncated power-law density profile. As mentioned before, these models have the advantage that they allow for a completely analytical characterisation of many dynamical properties, while the models still cover a wide range in density profiles.

This paper is organised as follows. In Section~{\ref{Basic.sec}} we discuss a number of general properties of the family of truncated power-law spheres that do not depend on the orbital structure. In Section~{\ref{DF.sec}} we determine the distribution function of the truncated power-law spheres with a TC orbital structure, and we investigate the consistency of these dynamical models as a function of the parameters $\beta_0$ and $\gamma$. In Sections~{\ref{DED.sec}} and~{\ref{Dispersions.sec}} we discuss the differential energy distribution and the velocity dispersions, respectively. In Section~{\ref{Discussion.sec}} we discuss the relevance of our set of models, and in particular focus on the new corners in the dynamical model parameter space that our family of models covers. We summarise our main findings in Section~{\ref{Summary.sec}}.


\section{Basic properties}
\label{Basic.sec}

\begin{figure*}
\centering
\includegraphics[width=0.95\textwidth]{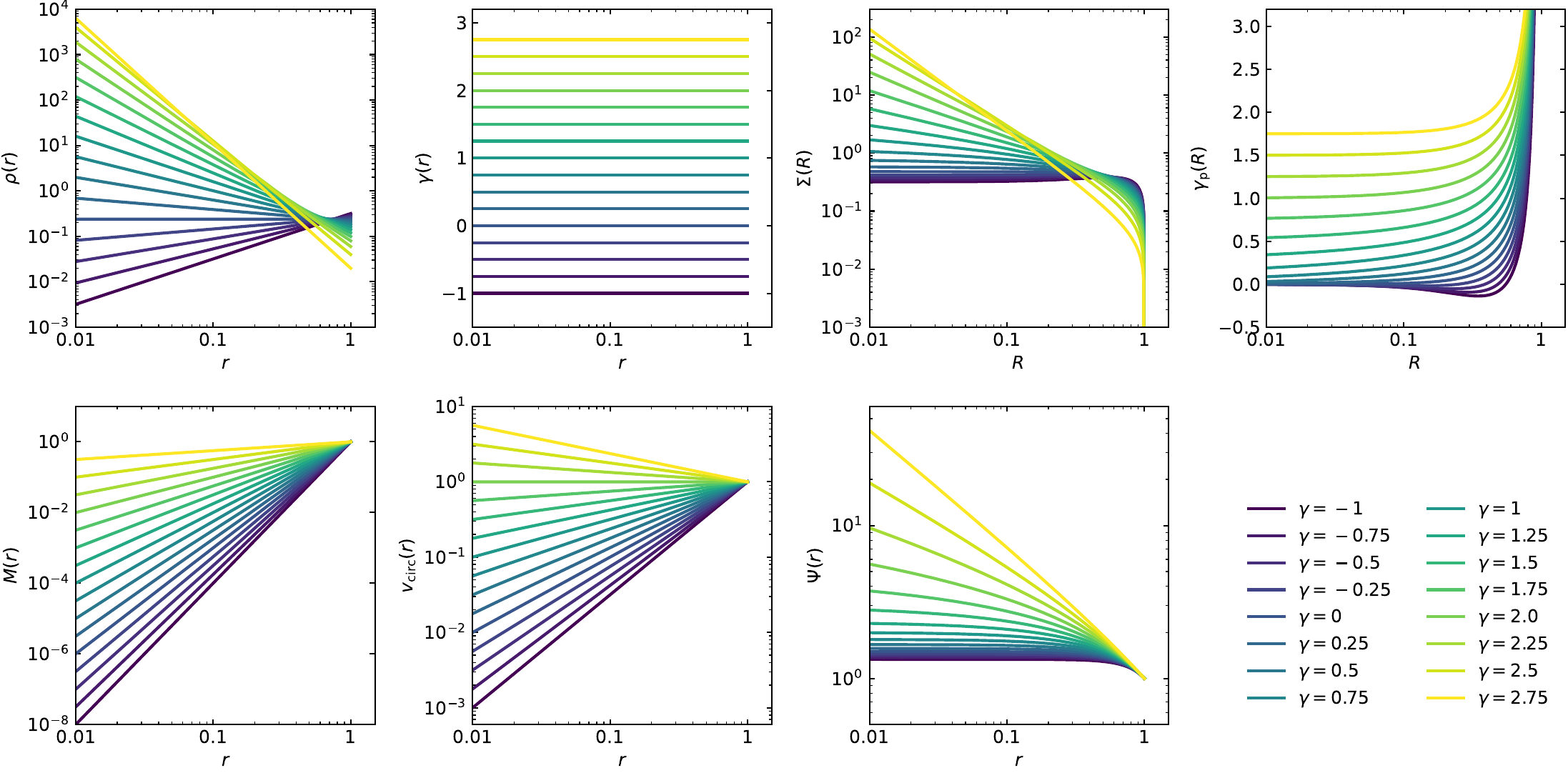}\hspace*{2em}
\caption{Basic properties of the family of truncated power-law spheres for different values of the power-law index $\gamma$. The different panels show the density, logarithmic density slope, surface density, logarithmic surface density slope, cumulative mass, circular velocity, and gravitational potential.}
\label{TPL-Basic.fig}
\end{figure*}

We define the family of truncated power-law spheres through the density profile
\begin{equation}
\rho(r) = \frac{3-\gamma}{4\pi}\, \frac{M}{\rT^3} \left(\frac{r}{\rT}\right)^{-\gamma}\,\Theta(\rT-r),
\end{equation}
with $M$ the total mass, $\rT$ the truncation radius, $\gamma$ the negative logarithmic density slope, and $\Theta(x)$ the Heaviside step function. The parameter $\gamma$ is limited to the range $\gamma<3$ to guarantee a finite total mass. To simplify the notations, we use dimensionless units with $G=M=\rT=1$, and when presenting expressions for radial profiles, we will only present the expressions for $r\leqslant1$. With these conventions, we can simply write
\begin{equation}
\rho(r) = \frac{3-\gamma}{4\pi}\, r^{-\gamma}.
\label{TPL-Density}
\end{equation}
The density profile is shown in Fig.\,{\ref{TPL-Basic.fig}}a for different values of $\gamma$. For $0<\gamma<3$ the density is a monotonically decreasing function of radius, for $\gamma=0$ we have a uniform density sphere \citepalias{2022MNRAS.512.2266B}, and for $\gamma<0$ the density is zero at the centre and it increases monotonically as a function of radius. In all cases, the negative logarithmic density slope,
\begin{equation}
\gamma(r) \equiv -\frac{\txd\log\rho}{\txd\log r}(r),
\end{equation}
is uniform up to the truncation radius, as shown in Fig.\,{\ref{TPL-Basic.fig}}b.

For such a simple density profile, the expression for the surface density profile is surprisingly complex. It can be expressed conveniently in terms of the complete and incomplete Beta functions,
\begin{equation}
\Sigma(R) = \frac{3-\gamma}{4\pi}\,R^{1-\gamma}
\left[{\text{B}}\left(\frac{\gamma-1}{2},\frac12\right)-{\text{B}}_{R^2}\!\left(\frac{\gamma-1}{2},\frac12\right)\right].
\label{TPL-SurfaceDensity}
\end{equation}
This expression reduces to simpler expressions involving only elementary functions for integer values of $\gamma$, for example
\begin{equation}
\Sigma(R)
=
\begin{cases}
\displaystyle 
\;\frac{3\,\sqrt{1-R^2}}{2\pi}
&{\text{if }}\gamma=0,
\\[1em]
\displaystyle 
\;\frac1\pi \arctanh\sqrt{1-R^2}
&{\text{if }}\gamma=1,
\\[1em]
\displaystyle 
\;\frac{1}{2\pi}\,\frac{\arccos R}{R}
&{\text{if }}\gamma=2.
\end{cases}
\end{equation}
As illustrated in Fig.\,{\ref{TPL-Basic.fig}}c, truncated power-law spheres with $\gamma<1$ have a finite central surface density, whereas models with $\gamma\geqslant1$ have a surface density cusp. The asymptotic behaviour for $R\to0$ is
\begin{equation}
\Sigma(R) \approx
\begin{cases}
\displaystyle
\;\frac{3-\gamma}{2\pi\,(1-\gamma)}
&{\text{if }}\gamma<1,
\\[1em]
\displaystyle 
\;\frac1\pi \ln\left(\frac2R\right)
&{\text{if }}\gamma=1,
\\[1em]
\displaystyle 
\;\frac{3-\gamma}{4\sqrt\pi}\,\frac{\Gamma\left(\frac{\gamma-1}{2}\right)}{\Gamma\left(\frac{\gamma}{2}\right)}\,R^{1-\gamma}
&{\text{if }}\gamma>1.
\end{cases}
\end{equation} 
The surface density profiles are monotonically decreasing for $\gamma\geqslant0$, whereas they show a local maximum for $\gamma<0$. This is most clearly seen when looking at the logarithmic surface density slope, 
\begin{equation}
\gamma_{\text{p}}(R) \equiv -\frac{\txd\log\Sigma}{\txd\log R}(R),
\end{equation}
shown in Fig.\,{\ref{TPL-Basic.fig}}d. Another obvious difference between the density and surface density profiles is the behaviour near the truncation radius. While the density is abruptly truncated at $r=1$, the surface brightness converges smoothly to zero for $R\to1$,
\begin{equation}
\Sigma(R) \approx \frac{3-\gamma}{\sqrt2\pi}\,\sqrt{1-R}.
\end{equation}

Like the density profile, the cumulative mass profile (Fig.\,{\ref{TPL-Basic.fig}}e) and the circular velocity curve (Fig.\,{\ref{TPL-Basic.fig}}f) are also simple power-law functions,
\begin{gather}
M(r) = r^{3-\gamma},
\label{TPL-Mass}
\\
v_{\text{circ}}(r) = r^{1-\gamma/2}.
\end{gather}
The gravitational potential (Fig.\,{\ref{TPL-Basic.fig}}g) is Keplerian for $r\geqslant 1$, whereas for $r\leqslant1$ we have
\begin{equation}
\Psi(r) = 
\begin{cases}
\displaystyle
\;\frac{1}{2-\gamma}\left[ (3-\gamma) - r^{2-\gamma}\right] & 
{\text{if }}\gamma\ne2, 
\\[1em]
\displaystyle
\;1 + \ln\left(\frac{1}{r}\right) & 
{\text{if }}\gamma=2.
\end{cases}
\label{TPL-Potential}
\end{equation}
Models with $\gamma<2$ have a finite potential well, whereas truncated power-law spheres with $2\leqslant \gamma < 3$ are characterised by an infinitely deep potential well,
\begin{equation}
\Psi_0 = 
\begin{cases}
\displaystyle
\;\frac{3-\gamma}{2-\gamma}
&{\text{if }}\gamma<2, 
\\[0.5em]
\displaystyle
\;\infty
&{\text{if }}\gamma\geqslant2.
\end{cases}
\end{equation}
The total potential energy is
\begin{equation}
W_{\text{tot}} 
=
\begin{cases}
\;-\dfrac{3-\gamma}{5-2\gamma} & 
{\text{if }}\gamma<\frac52,
 \\[0.5em]
\;-\infty &
{\text{if }}\gamma\geqslant\frac52.
\end{cases}
\label{TPL-Wtot}
\end{equation}
Interestingly, the total potential energy budget becomes infinitely large as soon as the logarithmic density slope is larger than $\tfrac52$. It is straightforward to check that any self-consistent model in which the density at small radii diverges as $r^{-5/2}$ or steeper has an infinite total potential energy budget. This is, for example, the case for those members of the family of $\gamma$--models with $\gamma\geqslant\tfrac52$ \citep{1993MNRAS.265..250D, 1994AJ....107..634T, 2004MNRAS.351...18B}.


\section{The distribution function}
\label{DF.sec}

In this Section we investigate the consistency of the family of truncated power-law spheres with a TC orbital structure with $\ra=\rT=1$. We will do this by explicitly calculating the phase-space distribution function and investigating for which values of the parameters $\gamma$ and $\beta_0$ this distribution function is positive over the entire phase space.

\subsection{Calculation of the distribution function}

In general, a TC orbital structure is characterised by an anisotropy structure of the form (\ref{TC-beta}). The distribution function can be written as 
\begin{equation}
f(\calE,L) 
= 
h(Q) \left(\frac{L^2}{2\ra^2}\right)^{-\beta_0},
\end{equation}
where $Q$ is a specific combination of the binding energy and angular momentum integrals of motion, 
\begin{equation}
Q = \calE + \frac{L^2}{2\ra^2}.
\label{Q-}
\end{equation}
For a given density profile, the inversion hence comes down to finding a univariate function, the function $h(Q)$. It can be obtained using an extended Eddington-like inversion formula,
\begin{equation}
h(Q) = \frac{(2\pi)^{-3/2}}{\Gamma(1-\lambda)\,\Gamma(1-\beta_0)}\,
\frac{\txd}{\txd Q}
\int_0^Q \frac{\txd^m \varrho}{\txd\Psi^m}\,\frac{\txd\Psi}{(Q-\Psi)^\lambda},
\label{cudd}
\end{equation}
where the function $\varrho(\Psi)$ is defined as
\begin{equation}
\varrho(\Psi) = \left[ \left(\frac{r}{\ra}\right)^{2\beta_0}\left(1-\frac{r^2}{\ra^2}\right)^{1-\beta_0} \rho(r)\right]_{r=r(\Psi)}
\label{varrho-gen}
\end{equation}
and where we have set
\begin{gather}
m = \floor\left(\tfrac32-\beta_0\right),\\
\lambda = \tfrac32-\beta_0-m.
\label{lambda}
\end{gather}
Expression~(\ref{cudd}) for $h(Q)$ is only valid when $\beta_0$ is not a half-integer number. If $\beta_0$ is a half-integer number, and we set $m=\frac32-\beta_0$, we have the simpler inversion formula
\begin{equation}
h(Q) = \frac{2^{m-5/2}}{\pi^2\,(2m-3)!!}\,\left[\frac{\txd^m\varrho}{\txd\Psi^m}\,\right]_{\Psi=Q}.
\end{equation}

For the family of truncated power-law spheres with $\ra=\rT=1$, we find the following expression for the function $\varrho(\Psi)$,
\begin{equation}
\varrho(\Psi) = \frac{3-\gamma}{4\pi}\,v(\Psi)^{2\beta_0-\gamma}\,\left[1-v(\Psi)^2\right]^{1-\beta_0}\,\Theta(\Psi-1).
\label{varrhoTC}
\end{equation}
where $v(\Psi)$ represents the radial coordinate written in terms of the gravitational potential,
\begin{equation}
v(\Psi) 
= 
\begin{cases}
\displaystyle
\;\left[(3-\gamma) - (2-\gamma)\,\Psi\right]^{\frac{1}{2-\gamma}}
& {\text{if }}\gamma\ne2,
\\[0.5em]
\displaystyle
\;{\text{e}}^{1-\Psi}
& {\text{if }}\gamma=2.
\end{cases}
\end{equation}
For the special cases where $\beta_0$ is a half-integer number, finding a closed expressions for the distribution function is relatively straightforward since we essentially only need to differentiate the augmented density function (\ref{varrhoTC}) several times. The resulting expressions for $h(Q)$ are algebraic functions of $Q$, multiplied by a Heaviside step function $\Theta(Q-1)$. For example, for $\beta_0=\tfrac12$ ($m=1$) we find
\begin{equation}
h(Q) = \frac{3-\gamma}{8\sqrt2\,\pi^3}\,\frac{(\gamma-1)+(2-\gamma)\,v(Q)^2}{v(Q)\,\sqrt{1-v(Q)^2}}\,\Theta(Q-1),
\label{h-beta05}
\end{equation}
and for $\beta_0=-\tfrac12$ ($m=2$) we have 
\begin{equation}
h(Q) = \frac{3\,(3-\gamma)}{4\sqrt2\,\pi^3}\,\frac{(1+\gamma) - \gamma\,v(Q)^2}{v(Q)^{5-\gamma}\,\sqrt{1-v(Q)^2}}\,\Theta(Q-1).
\label{h-betam05}
\end{equation}
For other half-integer values of $\beta_0$ similar formulae can be obtained. 

For the more general case when $\beta_0$ is not a half-integer number, the distribution function can be calculated by inserting the augmented density~(\ref{varrhoTC}) in Eq.~(\ref{cudd}) and subsequently evaluating the resulting integral. In general, it is not possible to derive a closed expression for the distribution function for all values of $\gamma$ and $\beta_0$. Explicit expressions can be obtained for selected values of $\gamma$ and $\beta_0$. In particular, for all integer values of $\beta_0$, the distribution function can be expressed in terms of special functions, although the expressions can be rather cumbersome.

Particularly interesting is the special case $\beta_0=0$, which corresponds to the TOM orbital structure. In this case, the explicit dependence on angular momentum drops out and the distribution function only depends on $Q$. After some calculation we find
\begin{equation}
h(Q) = 
\frac{3-\gamma}{4\sqrt2\,\pi^3} 
\left[ \frac{1}{\sqrt{Q-1}} + U(Q) \right] \Theta(Q-1)
\label{h-beta0}
\end{equation}
with 
\begin{equation}
U(Q) = 
\begin{cases}
\displaystyle
\;2\gamma\,\sqrt{Q-1}\,v(Q)^{\gamma-4}\,
\\[1em]
\displaystyle
\qquad\times\; {}_2F_1\left(\dfrac12,\dfrac{2}{2-\gamma};\dfrac32;1-\frac{1}{v(Q)^{2-\gamma}}\right)
&
{\text{if }}\gamma\ne2,
\\[1.2em]
\displaystyle
\;\sqrt{2\pi}\,{\text{e}}^{2(Q-1)}\,\erf\sqrt{\,2\,(Q-1)}
&
{\text{if }}\gamma=2.
\end{cases}
\end{equation}
This TOM distribution function can be written in terms of elementary functions for all values of $\gamma$ for which $2/(2-\gamma)$ is an integer or a half-integer number. In particular, for the uniform density sphere with $\gamma=0$, we immediately find 
\begin{equation}
h(Q) = \frac{3}{4\sqrt2\,\pi^3} \, \frac{\Theta(Q-1)}{\sqrt{Q-1}},
\end{equation}
in agreement with the results obtained in \citetalias{2022MNRAS.512.2266B}. 

\subsection{Radial orbital structure ($\beta_0=1$)}

In our analysis of the TC orbital structure we have, until now, implicitly assumed that $\beta_0<1$, such that the anisotropy of the system systematically changes from the value $\beta_0$ in the centre to completely tangential ($\beta=-\infty$) at the truncation radius. This situation changes when $\beta_0=1$: in this case the anisotropy profile (\ref{TC-beta}) becomes $\beta(r)=1$, and the models are completely radial at all radii. Dynamical models with a fully radial orbital structure have only orbits with $L=0$ populated, and are therefore characterised by a distribution function of the form
\begin{equation}
f(\calE,L) = h(\calE)\,\delta(L^2).
\label{df-radial}
\end{equation}
The function $h(\calE)$, which is now a function of the binding energy $\calE$ instead of $Q$, can again be determined using an Eddington-like inversion formula \citep{1984ApJ...286...27R, 2021isd..book.....C},
\begin{equation}
h(\calE) = \frac{1}{\sqrt2\,\pi^2}\,\frac{\txd}{\txd\calE} \int_0^\calE \frac{\varrho(\Psi)\,\txd\Psi}{\sqrt{\calE-\Psi}},
\end{equation}
with 
\begin{equation}
\varrho(\Psi) = \left[ r^2\,\rho(r) \right]_{r=r(\Psi)}.
\end{equation}

For the family of truncated power-law spheres, setting $\beta_0=1$ in expression~(\ref{varrhoTC}) yields
\begin{equation}
\varrho(\Psi) 
= 
\frac{3-\gamma}{4\pi}\,v(\Psi)^{2-\gamma}\,\Theta(\Psi-1).
\end{equation}
For all values of $\gamma$, this leads to the following expression for the function $h(\calE)$,
\begin{equation}
h(\calE)
=
\frac{3-\gamma}{4\sqrt2\,\pi^3}
\left[\frac{1}{\sqrt{\calE-1}}+ 2\,(\gamma-2)\sqrt{\calE-1} \right] \Theta(\calE-1).
\label{h-beta1}
\end{equation}

\subsection{General behaviour and consistency}

\begin{figure*}
\includegraphics[width=0.95\textwidth]{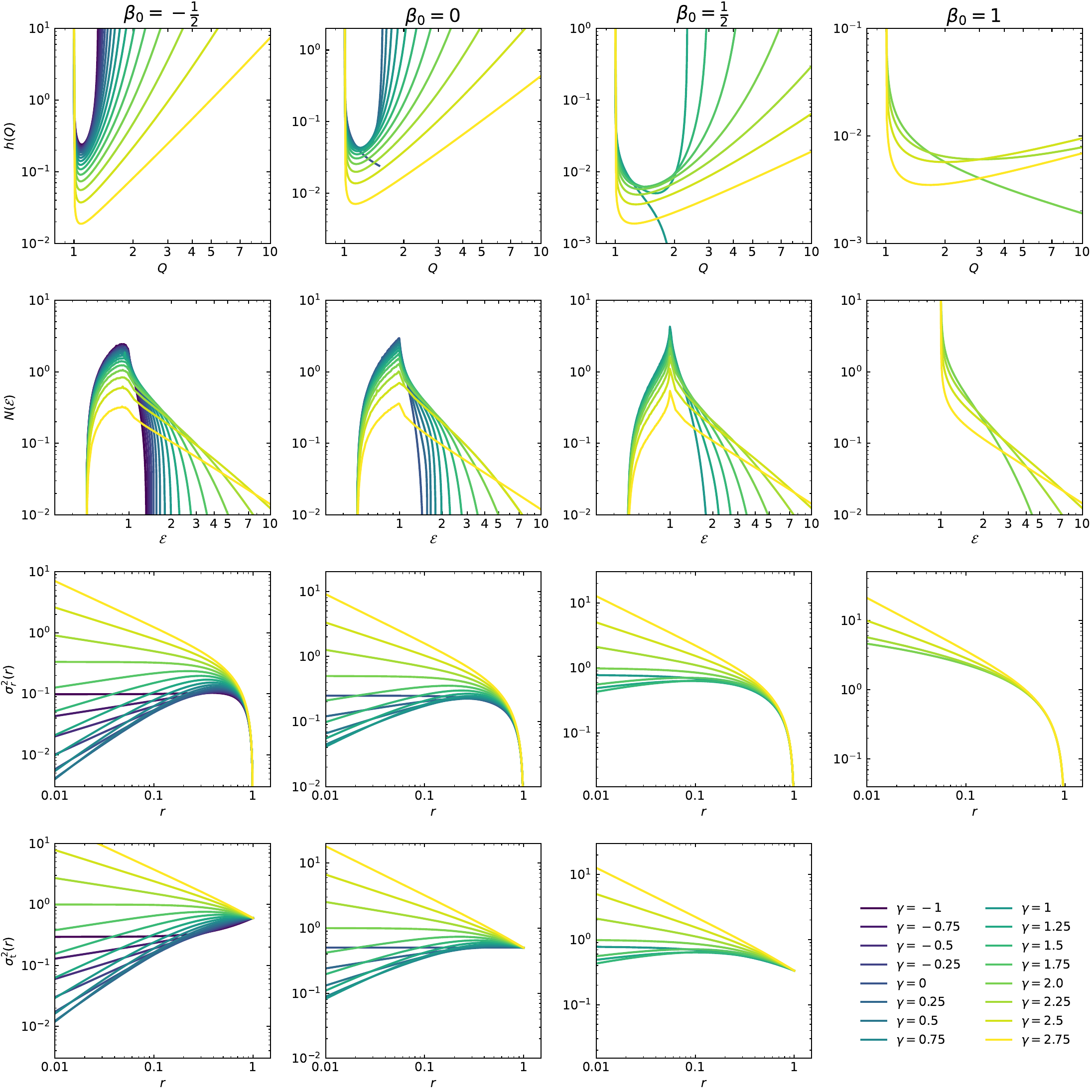}\hspace*{2em}
\caption{Dynamical properties of the family of truncated power-law spheres with a TC orbital structure, for different values of the power-law index $\gamma$ (different colours) and the central anisotropy $\beta_0$ (different columns). The different rows show the $Q$-dependent part of the distribution function, the differential energy distribution, the radial velocity dispersion profile, and the tangential velocity dispersion profile.}
\label{TPL-TC-all.fig}
\end{figure*}

On the top row of Fig.\,{\ref{TPL-TC-all.fig}} we plot the function $h(Q)$ for different values of $\gamma$ for the cases $\beta_0=-\tfrac12$, 0, $\tfrac12$ and 1, respectively. For each value of $\beta_0$, we show models corresponding to the same values of $\gamma$, but we only show those where the distribution function is positive over the entire phase space. It turns out that, for each value of $\beta_0$, the distribution function is always positive if and only if
\begin{equation}
\gamma \geqslant \gamma_{\text{min}}(\beta_0) = 2\beta_0.
\label{consistency}
\end{equation}
This can be easily checked for the cases $\beta_0=-\tfrac12$, 0, $\tfrac12$ and 1 using the explicit expressions~(\ref{h-betam05}), (\ref{h-beta0}), (\ref{h-beta05}), and (\ref{h-beta1}). 

The distribution functions shown on the top row of Fig.\,{\ref{TPL-TC-all.fig}} have a number of interesting characteristics. One obvious characteristic is that, for each choice of the model parameters, the distribution vanishes for all $Q<1$, which means that no orbits with $Q<1$ are populated. A second ubiquitous characteristic is that the functions $h(Q)$ diverge at the smallest binding energies. Concretely, we have an inverse square-root asymptotic behaviour for $Q\to1$,
\begin{equation}
h(Q) \propto \frac{1}{\sqrt{Q-1}}.
\end{equation}
At large $Q$, that is for $Q\to\Psi_0$, the asymptotic expansion is more complex and depends on both $\gamma$ and $\beta_0$. For all models with $\gamma>2\beta_0$, the function $h(Q)$ diverges at the high-$Q$ side,
\begin{equation}
h(Q) \propto
\begin{cases}
\displaystyle
\;(\Psi_0-Q)^{-\frac{6-8\beta_0-(1-2\beta_0)\gamma}{2\,(2-\gamma)}}
&
{\text{if }}2\beta_0<\gamma<2,
\\[1em]
\displaystyle
\;\txe^{2(1-\beta_0)\,Q}
&
{\text{if }}\gamma=2,
\\[1em]
\displaystyle
\;Q^{\frac{6-8\beta_0-(1-2\beta_0)\gamma}{2\,(\gamma-2)}}
&
{\text{if }}\gamma>2.
\end{cases}
\end{equation}
Combined with the inverse square-root divergence at the low-$Q$ side, this implies that the function $h(Q)$ has a non-monotonous U-shaped behaviour. The value of $Q$ at which the function $h(Q)$ reaches its minimum value depends on the parameters $\gamma$ and $\beta_0$. At fixed $\beta_0$, it decreases with increasing $\gamma$.

For the models with $\gamma=2\beta_0$, that is, the models at the boundary of the region in parameter space that still leads to consistent TC truncated power-law spheres, the asymptotic behaviour is slightly different, 
\begin{equation}
h(Q) \propto
\begin{cases}
\displaystyle
\;(\Psi_0-Q)^{-\frac{2-5\gamma+\gamma^2}{2\,(2-\gamma)}}
&
{\text{if }}\gamma=2\beta_0<0,
\\[1em]
\displaystyle
\;1 + (\Psi_0-Q)
&
{\text{if }}\gamma=2\beta_0=0,
\\[1em]
\displaystyle
\;(\Psi_0-Q)^{\gamma}
&
{\text{if }}0<\gamma=2\beta_0<2,
\\[1em]
\displaystyle
\;Q^{-\frac12}
&
{\text{if }}\gamma=2\beta_0=2.
\end{cases}
\end{equation}
For models with $\gamma=2\beta_0<0$, $h(Q)$ diverges as $Q$ approaches $\Psi_0$, like the models with $\gamma<2\beta_0$. For the model with $\gamma=2\beta_0=0$, that is, the uniform density sphere with the TOM orbital structure, $h(Q)$ converges to a finite value when $Q$ approaches $\Psi_0 = \tfrac32$. Finally, for models with $\gamma=2\beta_0>0$, $h(Q)$ is a monotonously decreasing function that converges to zero in the high-$Q$ limit.

The most important result is that the distribution function of the TC truncated power-law spheres is positive over the entire phase space if and only if the condition~(\ref{consistency}) is met. The existence of a lower limit on $\gamma$ for each $\beta_0$ is an expected result in light of the central density slope--anisotropy inequality \citep[CDSAI,][]{2006ApJ...642..752A}. This theorem states that any spherical dynamical model has to satisfy the inequality $\gamma(0) \geqslant 2\beta_0$. In fact, the CDSAI is a special case of the more general global density slope--anisotropy inequality \citep{2009MNRAS.393..179C, 2010MNRAS.408.1070C}, for which the validity conditions are still an open issue \citep{2011ApJ...726...80V, 2011MNRAS.413.2554A, 2011ApJ...736..151A, 2014MNRAS.442.3533B}. 

The CDSAI implies that not all truncated power-law spheres can be supported by all TC orbital structures. For models with a radially anisotropic centre ($\beta_0>0$), the range of compatible truncated power-law spheres is restricted to the larger values of $\gamma$ only. For the special case of $\beta_0=1$, we find $\gamma_{\text{min}}=2$, which means that only truncated power-law spheres with $\gamma\geqslant2$, that is, the ones with an infinitely deep potential well, can be supported by a completely radial orbital structure. This is in line with the well-known result that only models with density profiles that increase at least as fast as $r^{-2}$ at small radii can be consistent with completely radial orbits \citep{Bouvier1968, 1984ApJ...286...27R}. For $\beta_0=0$ the TC orbital structure reduces to the TOM orbital structure, and we find that only truncated power-law spheres with $\gamma\geqslant0$ can be supported by this orbital structure. For $\beta_0<0$, that is, models with a tangential central anisotropy, also truncated power-law spheres with negative values of $\gamma$ are possible. 

In general, the CDSAI is a necessary and not a sufficient condition for consistency: not all dynamical models that satisfy this inequality are consistent and hence physically meaningful. Simple counter-examples include the isotropic Einasto models with Einasto index $n<\tfrac12$ \citep{2022A&A...667A..47B}, or Osipkov--Merritt Plummer or Hernquist models with sufficiently small anisotropy radii \citep{1979PAZh....5...77O, 1985AJ.....90.1027M, 1995MNRAS.276.1131C, 2002A&A...393..485B}. For the present set of truncated power-law spheres with a TC orbital structure, however, we find that the CDSAI is both a sufficient and a necessary condition: all dynamical models allowed by the CDSAI are physically viable.


\section{The differential energy distribution}
\label{DED.sec}

The distribution function is the most fundamental quantity of any dynamical system: as the density distribution in phase space it contains all possible dynamical information. For our family of truncated power-law spheres, we have calculated explicit expressions for $f(\calE,L)$ for different orbital configurations. It is relatively hard to interpret the functional form of $f(\calE,L)$, however. 

A fundamental diagnostic of dynamical models that is easier to interpret is the differential energy distribution $N(\calE)$, that is, the distribution of mass as a function of $\calE$. It is a natural diagnostic for dynamical models that is easily calculated from $N$-body simulations \citep[e.g.,][]{1982MNRAS.201..939V, 2001ApJ...554.1268H, 2013MNRAS.431.3177D, 2020MNRAS.491.4591E}. It has been argued that the differential energy distribution is the most fundamental partitioning of an equilibrium stellar system \citep{1982MNRAS.200..951B, 2007LNP...729..297E, 2010ApJ...722..851H}.

\subsection{Calculation of the differential energy distribution}

For spherical dynamical models with an ergodic orbital structure, the differential energy distribution is relatively easy to calculate. For anisotropic models, this calculation is slightly more complicated. In general, it can be found as \citep{2008gady.book.....B}
\begin{equation}
N(\calE) = \int \txd\bfx \int \txd\bfv f(\bfx,\bfv)\,\delta\left(\Psi(\bfx)-\frac12|\bfv|^2-\calE\right).
\end{equation}
For spherical systems, this six-dimensional integral can be converted to a double integral \citep{2015MNRAS.454..576G, 2021A&A...653A.140B},
\begin{equation}
N(\calE) = 4\sqrt{2}\,\pi^2 \int_0^{r(\calE)} \txd r \int_0^{2r^2[\Psi(r)-\calE]}
\frac{f(\calE,L)\,\txd L^2}{\sqrt{\Psi(r)-\calE-\frac{L^2}{2r^2}}}.
\label{NE-gen}
\end{equation}
For our family of truncated power-law spheres with a TC orbital structure, there is no hope in finding an explicit expression for $N(\calE)$ for general values of $\gamma$ and $\beta_0$. Given explicit expressions for the distribution function, such as expressions (\ref{h-betam05}), (\ref{h-beta0}), or (\ref{h-beta05}), it can be evaluated numerically, however.

For the special case of radial orbit models, it is possible to obtain an explicit expression for $N(\calE)$. Indeed, as radial orbit models have a distribution function of the form (\ref{df-radial}), the inner integral in expression~(\ref{NE-gen}) is trivial, and we get
\begin{equation}
N(\calE) = 4\sqrt2\,\pi^2\, h(\calE) \int_0^{r(\calE)} \frac{\txd r}{\sqrt{\Psi(r)-\calE}}.
\end{equation}
For purely radial truncated power-law spheres, which all have $\gamma \geqslant 2$, we obtain the explicit expression
\begin{equation}
N(\calE)
=
\begin{cases}
\displaystyle
\;\frac{\sqrt\pi\,\txe^{-(\calE-1)}}{\sqrt{\calE-1}}\, \Theta(\calE-1)
&
{\text{if }}\gamma=2,
\\[1em]
\displaystyle
\;\frac{3-\gamma}{\sqrt\pi\,\sqrt{\gamma-2}}\,
\frac{\Gamma\left(\frac{\gamma}{2\gamma-4}\right)}{\Gamma\left(\frac{\gamma-1}{\gamma-2}\right)}\,
v(\calE)^{\gamma/2}
\\[1.5em]
\displaystyle
\qquad\times\,
\left[\frac{1+ 2\,(\gamma-2)\,(\calE-1)}{\sqrt{\calE-1}} \right] \Theta(\calE-1)
&
{\text{if }}\gamma>2.
\end{cases}
\label{NE-rad}
\end{equation}

\subsection{General behaviour}

In the second row of Fig.\,{\ref{TPL-TC-all.fig}} we show the differential energy distribution for the same set of truncated power-law spheres as considered in the top row. We have verified numerically that all the differential energy distributions are properly normalised, 
\begin{equation}
\int_0^\infty N(\calE)\,\txd\calE = 1.
\end{equation}
For all other models with $\beta_0<1$, the differential energy distributions have a similar general behaviour. A remarkable characteristic is that the minimum value for the differential energy distribution is $\calE_{\text{min}} = \tfrac12$, which corresponds to the binding energy of the circular orbit at the truncation radius,
\begin{equation}
\calE_{\text{min}} = \Psi(1) - \frac12\,v_{\text{circ}}^2(1) = \frac12.
\end{equation}
From zero at $\calE_{\text{min}}$ onwards, $N(\calE)$ gradually increases towards a finite maximum value at $\calE \sim 1$, and it subsequently gradually decreases towards zero at the $\calE=\Psi_0$. The exception to this general behaviour is the case $\beta_0=1$. In this case, all orbits are purely radial orbits and the lowest binding-energy orbit populated is the radial orbit with apocentre at the truncation radius, with has $\calE=1$. As a result, $N(\calE)$ is only non-zero for $1<\calE<\Psi_0$, and it diverges with an inverse square-root asymptotic behaviour for $\calE\to1$,
\begin{equation}
N(\calE) 
\propto
\frac{1}{\sqrt{\calE-1}}.
\end{equation}

At a fixed value of $\beta_0$, it can be noted that the differential energy distribution becomes progressively more skewed to larger values of $\calE$ as $\gamma$ increases. This systematic change can be quantified by the mean binding energy, or equivalently, the total integrated binding energy,
\begin{equation}
B_{\text{tot}} \equiv M_{\text{tot}} \langle\calE\rangle = \int_0^\infty N(\calE)\,\calE\,\txd\calE.
\label{Btotdef}
\end{equation}
For $\gamma\geqslant\tfrac52$, $N(\calE)$ decreases asymptotically slower than $\calE^{-2}$, such that the total integrated binding energy is infinitely large. For smaller values of $\gamma$, we find that $B_{\text{tot}}$ is finite and that its value increases with increasing $\gamma$, as expected, but that it does not depend on $\beta_0$,
\begin{equation}
B_{\text{tot}} =
\begin{cases}
\; \dfrac{3\,(3-\gamma)}{2\,(5-2\gamma)}
&{\text{if }}2\leqslant\gamma<\tfrac52,
\\[0.5em]
\;+\infty 
&{\text{if }}\gamma\geqslant\tfrac52.
\end{cases}
\label{TPL-Btot}
\end{equation}
This result can easily be checked for the special case $\beta_0 = 1$ by inserting the explicit expression~(\ref{NE-rad}) for the differential energy distribution in the integral (\ref{Btotdef}). Comparing the total integrated binding energy to the total potential energy (\ref{TPL-Wtot}) of the truncated power-law spheres, we have the relation
\begin{equation}
B_{\text{tot}} = -\frac32\,W_{\text{tot}}.
\end{equation}
This relation is generally valid for all self-consistent dynamical models \citep{2021A&A...653A.140B}.

Comparing curves for a fixed value of $\gamma$ but with different $\beta_0$, the discussion above already indicates that the differential energy distributions have the same normalisation and the same mean value. There is still a systematic change as a function of $\beta_0$ however: for small values of $\beta_0$, the differential energy distributions are broader, whereas for large values they are more peaked around the mean binding energy. Also this is a general characteristic of differential energy distributions of spherical dynamical models: the width of the differential energy distribution varies systematically with the global anisotropy. Radially anisotropic models tend to prefer more average binding energies, whereas models with a more tangential orbital distribution slightly favour more extreme binding energies \citep{2021A&A...653A.140B}.


\section{The velocity dispersions}
\label{Dispersions.sec}

\subsection{Radial velocity dispersion}

The radial velocity dispersion profile can be calculated by multiplying the distribution function with $v_r^2$ and integrating over velocity space. A simpler way, however, is to use the relation
\begin{equation}
\sigma_r^2(r) = \frac{1}{\varrho(r)} \int_r^1 \frac{\varrho(u)\,GM(u)\,\txd u}{u^2}.
\label{Jeans}
\end{equation}
Inserting expressions~(\ref{TPL-Density}), (\ref{TPL-Mass}), and (\ref{varrhoTC}) we find
\begin{equation}
\sigma_r^2(r) = \frac{r^{\gamma-2\beta_0}}{2\,(1-r^2)^{1-\beta_0}}\,
\Bigl[ {\text{B}} - {\text{B}}_{r^2}(1+\beta_0-\gamma,2-\beta_0)\Bigr],
\label{sigmar2-TC}
\end{equation}
where we have used the shorthand notation 
\begin{equation}
\txB = \txB(1+\beta_0-\gamma,2-\beta_0).
\end{equation}
If $\beta_0$ or $\beta_0-\gamma$ is an integer value, the velocity dispersions can be written in terms of elementary functions. In particular, for the special case of the TOM orbital structure ($\beta_0=0$) we find
\begin{equation}
\sigma_r^2(r) = 
\begin{cases}
\displaystyle
\;\frac{r^{\gamma} - (2-\gamma)\,r^{2-\gamma} + (1-\gamma)\,r^{4-\gamma}}{2\,(2-\gamma)\,(1-\gamma)\,(1-r^2)}
& {\text{if }}\gamma\ne1,2,
\\[1em]
\displaystyle
\;\frac{r}{1-r^2}\ln\left(\frac{1}{r}\right) - \frac{r}{2}
& {\text{if }}\gamma=1,
\\[1em]
\displaystyle
\;\frac12 - \frac{r^2}{1-r^2}\ln\left(\frac{1}{r}\right)
& {\text{if }}\gamma=2.
\end{cases}
\label{RadialDispersionTOM}
\end{equation}
and for the radial orbit models ($\beta_0=1$)
\begin{equation}
\sigma_r^2(r) 
=
\begin{cases}
\displaystyle
\;\ln\left(\frac1r\right)
&
{\text{if }}\gamma=2,
 \\[1em]
\displaystyle
\;\frac{r^{2-\gamma}-r^{\gamma-2}}{2\,(\gamma-2)}
&
{\text{if }}\gamma>2.
\end{cases}
\end{equation}
On the third row of Fig.\,{\ref{TPL-TC-all.fig}} we plot the radial velocity dispersion profiles for our set of truncated power-law spheres. The behaviour at small radii depends strongly on the values of $\gamma$ and $\beta_0$.  For $\beta_0<1$, we obtain the following asymptotic expression,
\begin{equation}
\sigma_r^2(r) \approx
\begin{cases}
\displaystyle
\;\frac12\,\txB+ \frac12\,(1-\beta_0)\,\txB\,r^2
&
{\text{if }}\gamma=2\beta_0{\text{ and }}\beta_0<0,
\\[1em]
\displaystyle
\;\frac14 - \frac{r^2}{4}
&
{\text{if }}\gamma=2\beta_0{\text{ and }}\beta_0=0,
\\[1em]
\displaystyle
\;\frac12\,\txB - \frac{r^{2\,(1-\beta_0)}}{2-2\beta_0}
&
{\text{if }}\gamma=2\beta_0{\text{ and }}\beta_0>0,
\\[1em]
\displaystyle
\;\frac12\,\txB\,r^{\gamma-2\beta_0}
&
{\text{if }}2\beta_0 < \gamma<1+\beta_0
\\[1em]
\displaystyle
\;r^{1-\beta_0} \ln\left(\frac{1}{r}\right)
&
{\text{if }}\gamma=1+\beta_0,
\\[1em]
\displaystyle
\;\frac{r^{2-\gamma}}{2\,(\gamma-\beta_0-1)}
&
{\text{if }}1+\beta_0<\gamma<2,
\\[1em]
\displaystyle
\;\frac{1}{2\,(1-\beta_0)} + \frac{r^2}{2\beta_0}
&
{\text{if }}\gamma=2{\text{ and }}\beta_0<0,
\\[1em]
\displaystyle
\;\frac{1}{2} - r^2 \ln\left(\frac{1}{r}\right)
&
{\text{if }}\gamma=2{\text{ and }}\beta_0=0,
\\[1em]
\displaystyle
\;\frac{1}{2\,(1-\beta_0)} + \frac12\,\txB\,r^{2-2\beta_0}
&
{\text{if }}\gamma=2{\text{ and }}\beta_0>0,
\\[1em]
\displaystyle
\;\frac{1}{2\,(\gamma-\beta_0-1)\,r^{\gamma-2}}
&
{\text{if }}\gamma>2.
\end{cases}
\end{equation}
For each value of $\beta_0<1$, there are two values of $\gamma$ for which the radial dispersion profile converges to a finite non-zero value at small radii: the minimum value $\gamma=\gamma_{\text{min}}=2\beta_0$ and the case $\gamma=2$ corresponding to the truncated SIS. For all values of $\gamma$ between these two cases, the radial velocity dispersion shows a central cavity, that is, it drops to zero in the centre. The strength of this central cavity increases when $\gamma$ increases beyond the minimum value $\gamma_{\text{min}}$ until it reaches the strongest depression for $\gamma = 1+\beta_0$, and subsequently the depression becomes gradually weaker until it disappears for $\gamma=2$. For all $\gamma>2$, the radial dispersion profile diverges at small radii with a slope that depends on $\gamma$ but not on $\beta_0$. 

For the special case of radial orbit models, the two special values $\gamma=2\beta_0$ and $\gamma=2$ coincide, and we have as asymptotic behaviour
\begin{equation}
\sigma_r^2(r) \approx
\begin{cases}
\displaystyle
\;\ln\left(\frac{1}{r}\right)
&
{\text{if }}\gamma=2,
\\[1em]
\displaystyle
\;\frac{1}{2\,(\gamma-2)\,r^{\gamma-2}}
&
{\text{if }}\gamma>2.
\end{cases}
\end{equation}

At radii close to the truncation radius, $r\lesssim1$, the radial dispersion profiles all converge to zero in a way that is independent of the value of $\gamma$,
\begin{equation}
\sigma_r^2(r) \approx \frac{1-r}{2-\beta_0}.
\end{equation}
This is expected since the only orbits that contribute to the density at the truncated radius are those with exactly the truncation radius as the apocentre. By definition, all orbits have $v_r=0$ at their apocentre, and therefore the radial velocity dispersion drops to zero at $r=1$.

\subsection{Tangential velocity dispersion}

Given the anisotropy profile (\ref{TC-beta}) of the TC orbital structure, we find for the tangential velocity dispersion 
\begin{equation}
\sigma_\txt^2(r) = 2\left[1-\beta(r)\right] \sigma_r^2(r) = \frac{2\,(1-\beta_0)\,\sigma_r^2(r)}{1-r^2}.
\label{sigmat2-TC}
\end{equation}
The tangential velocity dispersion profiles are shown on the bottom row of Fig.\,{\ref{TPL-TC-all.fig}}. At small radii, we logically have the same asymptotic behaviour as the radial dispersion profiles, apart from an additional scaling factor $2\,(1-\beta_0)$. The interesting difference is found at radii close to the truncation radius: while the radial dispersion profiles all converge to zero for $r\to1$, the tangential dispersion profiles all converge to a non-zero value that is independent of the value of $\gamma$,
\begin{equation}
\sigma_\txt^2(1) 
=
\frac{1-\beta_0}{2-\beta_0}.
\end{equation} 
Only for completely radial orbit models, the tangential dispersion also vanishes at the truncation radius; in fact, for radial orbit models, the tangential dispersion is identically zero at all radii.
 
\subsection{The kinetic energy budget and stability issues}

\begin{figure}
\includegraphics[width=0.93\columnwidth]{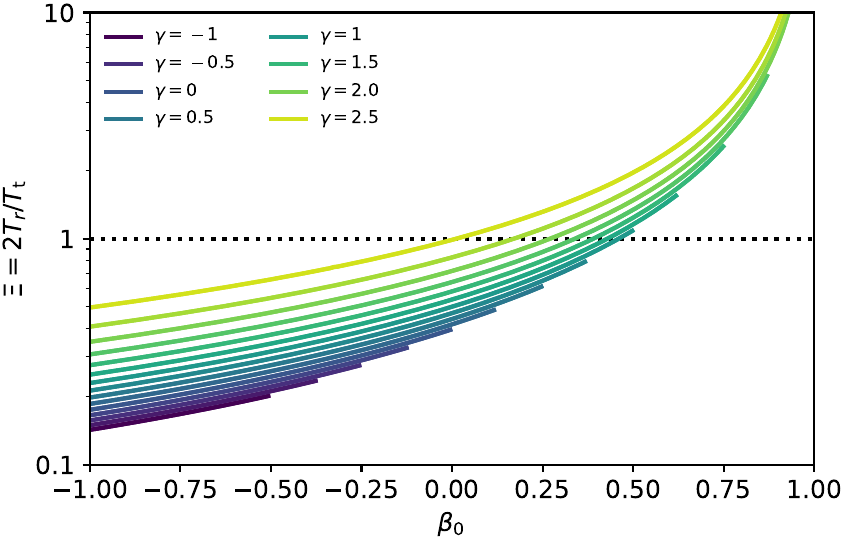}\hspace*{2em}
\caption{The ratio $\Xi = 2T_r/T_\txt$ as a function of the central anisotropy $\beta_0$ for the family of truncated power-law spheres, for different values of the power-law index $\gamma$. The dotted horizontal line indicates equipartition between the total radial and tangential kinetic energy.}
\label{TPL-EnergyXi.fig}
\end{figure}

Given the expressions~(\ref{sigmar2-TC}) and (\ref{sigmat2-TC}) for the radial and tangential dispersions, we can calculate the total kinetic energy. For arbitrary values of $\gamma$ and $\beta_0$ we find that
\begin{equation}
T_{\text{tot}} = 
\begin{cases}
\; \dfrac{3-\gamma}{2\,(5-2\gamma)}
&{\text{if }}\gamma<\tfrac52,
\\[1em]
\;+\infty 
&{\text{if }}\gamma\geqslant\tfrac52.
\end{cases}
\label{Ttot}
\end{equation}
Comparing this with Eqs.~(\ref{TPL-Wtot}) and (\ref{TPL-Btot}), we obtain
\begin{equation}
B_{\text{tot}} = 3\,T_{\text{tot}} = -\frac32\,W_{\text{tot}},
\end{equation}
which is the extension of the virial theorem that should be satisfied by all self-gravitating equilibrium dynamical models \citep{2021A&A...653A.140B}.

Apart from the global kinetic energy budget, we can also consider the separate contributions of the total radial and tangential kinetic energies. A general expression for arbitrary values of $\gamma$ and $\beta_0$ cannot be obtained, but these quantities are easily obtained numerically. In Fig.~{\ref{TPL-EnergyXi.fig}} we show the global anisotropy indicator $\Xi=2T_r/T_\txt$ as a function of $\beta_0$ for different values of $\gamma$. This plot shows that $\Xi$ is an increasing function of both $\beta_0$ and $\gamma$. The increase of $\Xi$ with increasing $\beta_0$ is logical, as for every fixed $\gamma$, models with increasing $\beta_0$ are increasingly radially anisotropic at every radius. The increase with increasing $\gamma$ stems from the fact that models with increasing $\gamma$ are more centrally concentrated, and models with a TC orbital structure become systematically more tangential at large radii. 

Only a few models in our two-parameter family of TC truncated power-law spheres can be characterised as globally radially anisotropic ($\Xi>1$). These few models have $\beta_0>0$ and a large central density concentration (large values of $\gamma$). The shallowest model that still reaches global radial anisotropy has $\gamma\approx0.9230$ and the maximum central anisotropy for this density profile ($\beta_0=\tfrac12\gamma \approx 0.4615$). Slightly surprising is that for $\beta_0=0$, the equipartition between the total radial and tangential kinetic energy is already reached for $\gamma\to\tfrac52$: these models are strongly centrally concentrated, but still tangentially anisotropic $(\beta<0)$ at all radii. 

The ratio $\Xi = 2T_r/T_\txt$ is commonly used as an indicator of the stability of spherical models against radial orbit instabilities (ROI). Based on a study of different families of radially anistropic models, \citet{1981SvA....25..533P} and \citet{1984sv...bookQ....F} argued that the ROI starts to kick in for $\Xi > \Xi_{\text{crit}} = 1.7 \pm 0.25$. This implies that models with $\Xi<\Xi_{\text{crit}}$ are stable against the ROI, which has become known as the Fridman--Polyachenko--Shukhman stability indicator. Stability studies of different families of dynamical models, using both linear stability analysis and N-body simulations, revealed a slightly wider scatter for $\Xi_{\text{crit}}$, with values roughly in the range $1.4 < \Xi_{\text{crit}} < 3$ \citep[e.g.,][]{1988ApJ...328...93D, 1991MNRAS.248..494S, 1991ApJ...368...66W, 1994ApJ...434...94B, 1997ApJ...490..136M, 2009ApJ...704..372B, 2011MNRAS.416.1836P}. 

Looking at Fig.~{\ref{TPL-EnergyXi.fig}}, we see that the vast majority of the TC truncated power-law spheres satisfy the Fridman--Polyachenko--Shukhman stability indicator. This is obviously not surprising given that the orbital structure varies from $\beta_0$ in the centre to completely tangentially anisotropic at the truncation radius. Only a small number of models with very steep density profiles and strong central radial anisotropy is possibly unstable against the ROI. Whether or not this actually happens requires a more detailed stability investigation, which is beyond the scope of this paper.


\section{Discussion}
\label{Discussion.sec}

\begin{figure}
\includegraphics[width=0.88\columnwidth]{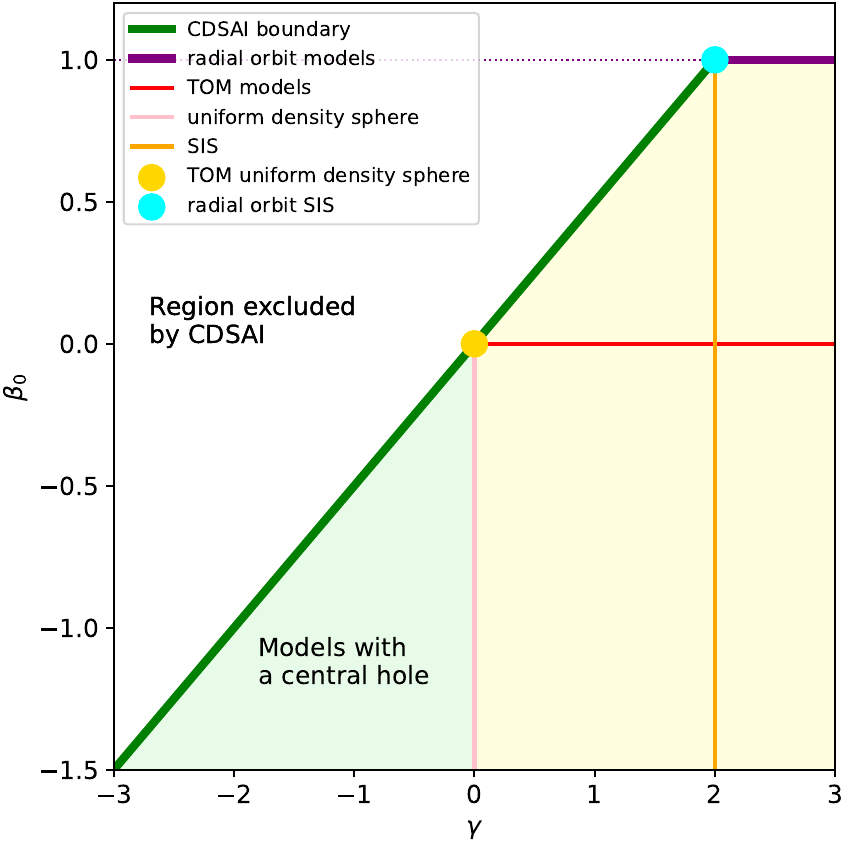}\hspace*{2em}
\caption{The $(\gamma,\beta_0)$--parameter space of the family of truncated power-law spheres with a TC orbital structure.}
\label{ParameterSpace.fig}
\end{figure}

The main goal of this paper was substantially enlarging the suite of simple analytical dynamical models with a finite extent. Looking at self-consistent dynamical models that have both an analytical density profile with a finite extent and an analytical distribution function, two examples come to mind: the uniform density sphere with TOM orbital structure (\citealt{1974SvA....17..460P, 1979PAZh....5...77O, 2021Ap.....64..219B}; \citetalias{2022MNRAS.512.2266B}) and the truncated SIS with a completely radial orbital structure \citep{1984sv...bookQ....F}. Both of these dynamical models are special cases of the broad two-parameter family of truncated power-law spheres with TC orbital structure that we present here, and as shown in Fig.\,{\ref{ParameterSpace.fig}}, they are at the border of the allowed $(\gamma,\beta_0)$--parameter space. In general, we thus significantly extent the available parameter space of analytical dynamical models with finite extent.

In the following two subsections we focus on two special regions in the $(\gamma,\beta_0)$--parameter space that deserve some special attention.

\subsection{The truncated SIS and other purely radial orbit models}

Throughout this paper, one specific model stood out among the family of truncated power-law sphere: the truncated SIS corresponding to $\gamma=2$. This model is special in that it forms the bridge between models with a finite potential well ($\gamma<2$) and models with an infinitely deep potential well ($\gamma>2$). Regardless of the central anisotropy, the truncated SIS also has radial and tangential velocity dispersion profiles that always converge to finite non-zero values at small radii.  

The truncated SIS is also particular as the model with shallowest cusp that can be supported by any TC orbital structure. In particular, it corresponds to lowest value of $\gamma$ that allows for a completely radial orbital structure. The fact that radial orbit models require a density profile that increases at least as fast as $r^{-2}$ at small radii has been known for decades \citep{Bouvier1968, 1984ApJ...286...27R} and is a special case of the CDSAI \citep{2006ApJ...642..752A}. The interesting aspect of the purely radial truncated SIS is that all of its dynamical properties are extremely simple,
\begin{gather}
\rho(r) = \frac{1}{4\pi\,r^2},
\\
\Psi(r) = 1- \ln r,
\\
f(\calE,L) = \frac{1}{4\sqrt2\,\pi^3}\,\frac{\Theta(\calE-1)}{\sqrt{\calE-1}}\,\delta(L^2),
\\
N(\calE) = \frac{\sqrt\pi\,\txe^{-(\calE-1)}}{\sqrt{\calE-1}}\,\Theta(\calE-1),
\\
\sigma_r^2(r) = -\ln r.
\end{gather}
The purely radial SIS was first presented by \citet{1984sv...bookQ....F}, and was probably the first self-consistent model with a finite mass and a purely radial orbital structure.\footnote{\citet{1987gady.book.....B} note that the truncated SIS was the only analytical radial orbit model known to them at that time.} To the best of our knowledge, only two other completely analytical radial orbit models have been presented in the literature: the radial Jaffe model \citep{1985MNRAS.214P..25M} and the radial Dehnen model with $\gamma=\tfrac52$ \citep{2004MNRAS.351...18B}. 

In Section~{\ref{DF.sec}} we showed that all truncated power-law spheres with $\gamma\geqslant2$ can be supported by a purely radial orbital structure. Not only for the truncated SIS but also for all truncated power-law spheres with $\gamma>2$ the distribution function and the differential energy distribution can be expressed as simple analytical functions. Another particularly simple model is the radial truncated power-law sphere with $\gamma=\tfrac52$, for which we readily obtain
\begin{gather}
\rho(r) = \frac{1}{8\pi\,r^{5/2}},
\\
\Psi(r) = \frac{2}{\sqrt{r}}-1,
\\
f(\calE,L) = \frac{1}{8\sqrt2\,\pi^3}\,\frac{\calE\,\Theta(\calE-1)}{\sqrt{\calE-1}}\,\delta(L^2),
\\
N(\calE) = \frac{3\calE\,\Theta(\calE-1)}{2\,(1+\calE)^{5/2}\sqrt{\calE-1}},
\\
\sigma_r^2(r) = \frac{1}{\sqrt{r}}-\sqrt{r}.
\end{gather}
This set of radial orbit models forms a significant extension of the limited set of radial orbit models presented up to now.

\subsection{Models with a central hole}

Nearly all of the models used in dynamics studies have a density profile that decreases for increasing radius, or equivalently, with $\gamma(r)>0$. The truncated power-law spheres with $\gamma>0$ obviously share this characteristic, and they are the main focus of the analysis in this paper. The uniform density sphere, corresponding to $\gamma=0$, seems a natural boundary case. However, from a purely mathematical or technical point of view, there is no formal need to stop at $\gamma=0$, and we can also consider truncated power-law spheres with negative values of $\gamma$. These models have a central hole in their density profile, i.e., $\rho(r)\to0$ for $r\to0$, that gradually grows stronger as $\gamma$ grows more negative. In the limit $\gamma\to-\infty$, the model reduces to an infinitely thin shell at the break radius. 

The entire analysis applied in Sections~{\ref{Basic.sec}} to {\ref{Dispersions.sec}} is insensitive to the sign of $\gamma$, meaning that none of the formulae derived do assume, implicitly or explicitly, that $\gamma\geqslant0$. For example, the expressions~({\ref{TPL-SurfaceDensity}}), ({\ref{TPL-Mass}}), and ({\ref{TPL-Potential}}) for the surface density, cumulative mass, and gravitational potential, respectively, are perfectly valid for negative $\gamma$ as well.

One of the main conclusions of our investigation is that, for our family of truncated power-law sphere with a TC orbital structure, the CDSAI is a necessary and sufficient criterion for consistency. In other words, all truncated power-law spheres with logarithmic density slope $\gamma$ can be supported by the TC orbital structure with central anisotropy $\beta_0$ if and only if $\gamma\geqslant2\beta_0$. For models with a tangentially anisotropic centre ($\beta_0<0$), we obtain the interesting result that not only all standard truncated power-law spheres with $\gamma\geqslant0$ are compatible, but also a range of truncated power-law spheres with negative $\gamma$. For example, setting $\beta_0=-\tfrac12$ and $\gamma=-1$, we find a very simple self-consistent dynamical model for a sphere in which the density simply increases linearly with radius. For this model, most of the dynamical properties can be expressed analytically, for example
\begin{gather}
\rho(r) = \frac{r}{\pi},
\\
\Psi(r) = \frac{4-r^3}{3},
\\
f(\calE,L) = \frac{3}{2\pi^3}\,\frac{L\,\Theta(Q-1)}{(4-3Q)^{4/3} \sqrt{1-(4-3Q)^{2/3}}},
\\
\sigma_r^2(r) = \frac{1}{48}\left[\frac{r\,(1-4r^2)\,(3-2r^2)}{1-r^2} + \frac{3\arccos r}{(1-r^2)^{3/2}}\right].
\end{gather}
To the best of our knowledge, this is the first non-trivial self-consistent distribution function for a model with a central hole in the density distribution. While this toy model obviously does not resemble a realistic stellar system or dark matter halo, it is inspiring to see how far we can stretch our set of models. Moreover, models as these can be used as a challenging case to test numerical orbit integrators or to study the development of instabilities in stellar systems.


\section{Summary}
\label{Summary.sec}

In this work, we investigate the self-consistency and dynamical properties of power-law mass distributions with a finite extent. These truncated power-law spheres form an important family of spherical dynamical models, which includes the point mass ($\gamma \to 3$), the uniform density sphere ($\gamma = 0$), and models with a central density hole ($\gamma < 0$). These models are very useful as a starting point for complex numerical modelling, as many structural and dynamical properties can be calculated analytically. However, we found that these derived quantities can quickly become quite complex, see Section~{\ref{Basic.sec}}.

Analytical models to describe the dynamical structure of gravitationally bound systems are mostly infinite in extent, while true astrophysical systems are finite. The default method to generate dynamical models with a finite extent is to apply an energy truncation to the distribution function of models with infinite extent. This, unfortunately, does not lead to models in which the density, let alone other dynamical properties, can be calculated analytically. In this sequence of papers, we follow an alternative path, holding on to a preset finite density profile. In \citetalias{2022MNRAS.512.2266B} we investigated the detailed dynamical structure of the uniform density sphere, probably the simplest model with a finite extent. In \citetalias{2023MNRAS.519.6065B}, we formulated a consistency hypothesis that states that dynamical models with a finite extent can often be supported by a TOM orbital structure, but not by an ergodic orbital structure. In this third paper, we generalise both results by looking at the more general TC orbital structure, applied to the broad family of truncated power-law spheres.

In Section~{\ref{DF.sec}}, we investigate the self-consistency of TC truncated power-law spheres as a function of the negative logarithmic density slope $\gamma$ and the central anisotropy $\beta_0$. This means that we calculate the phase-space distribution function, and assess if the resulting orbit configuration is physical. The general procedure for calculating the DF is outlined in Section~{\ref{DF.sec}}, which can be followed for all $\gamma$ and $\beta_0$. In addition, we provide closed analytical expressions for selected parameter combinations of $\gamma$ and $\beta_0$. We find that truncated power-law spheres can be supported by a TC orbital structure when the negative logarithmic density slope $\gamma$ is larger or equal than two times the central anisotropy $\beta_0$. This means that these models can only have a radially anisotropic centre ($\beta_0 > 0$) when the density profile is a decreasing function of radius ($\gamma > 0$), while models with some tangential central anisotropy ($\beta_0 < 0$) can support a central density hole ($\gamma < 0$). Finally, truncated power-law spheres can be supported by a TOM orbital structure ($\beta_0 = 0$) when $\gamma \geqslant 0$. This result was interpreted in the context of the central density slope-anisotropy inequality, which forms a sufficient condition for self-consistency in case of TC truncated power-law spheres.

In Section~{\ref{DED.sec}}, the orbit occupancy of self-consistent truncated power-law spheres was interpreted in terms of the differential energy distribution, which was evaluated numerically for $\beta_0 < 1$. For $\beta_0=1$, we provide an analytical distribution function, with binding energies between 1 and $\Psi_0$ (the central potential), which diverges at $\calE \to 1$. In the general case of non-radial anisotropy profiles ($\beta_0 < 0$), binding energies are distributed between $\tfrac12$ (i.e. the binding energy of a circular orbit at the truncation radius) and $\Psi_0$, and peak at $\calE \approx 1$. The differential energy distribution becomes more skewed to larger values of $\calE$ with increasing $\gamma$, and is more narrowly peaked around the mean binding energy with increasing $\beta_0$ (which is an expected result for spherical dynamical models).

Finally, we provide radial and the tangential velocity dispersion profiles in Section~{\ref{Dispersions.sec}}, which can be calculated analytically for all self-consistent truncated power-law models. In addition, we calculate the global anisotropy, and find that the relative contribution of the total radial kinetic energy increases for increasing $\gamma$ and $\beta_0$, as expected from the TC anisotropy profile. Together with the structural properties presented in Section~{\ref{Basic.sec}}, these velocity dispersion profiles will be useful for a comparison to more complex numerical modelling results.

This work significantly enlarges the available suite of self-consistent dynamical models with a finite extent. Our truncated power-law spheres have a simple analytical density profile for which the structural and dynamical properties can be derived analytically, which is not true for finite models with an energy truncation to the distribution function. In fact, previous examples of self-consistent finite models such as the uniform density sphere and the truncated SIS are just special cases of the two-parameter family of TC truncated power-law spheres, which cover a much broader parameter space including a new set of radial orbit models, and a non-trivial family of models with a central density hole. The dynamical models in our two-parameter family are idealised systems and therefore do not directly represent real dynamical structures. They can, however, be useful as first-order approximations for truncated dynamical systems such as global clusters or dark matter haloes, and they can serve as theoretical laboratories to test numerical orbit integrators or study the development of instabilities in systems with a finite extent.

\section*{Acknowledgements}

BVM acknowledges the financial support from the Fund for Scientific Research Flanders (FWO-Vlaanderen, PhD Fellowship Grant 11H2123N). We thank the anonymous reviewer for the insightful comments and suggestions.

\section*{Data availability}

No astronomical data were used in this research. The data generated and the plotting routines will be shared on reasonable request to the corresponding author.

\bibliographystyle{mnras}
\bibliography{TPL_bib}

\bsp	
\label{lastpage}
\end{document}